\documentclass[lettersize,journal]{IEEEtran}
\usepackage{amsmath,amsfonts}
\usepackage{algorithmic}
\usepackage{algorithm}
\usepackage{array}
\usepackage[caption=false,font=normalsize,labelfont=sf,textfont=sf]{subfig}
\usepackage{textcomp}
\usepackage{stfloats}
\usepackage{url}
\usepackage{verbatim}
\usepackage{graphicx}
\usepackage{cite}
\usepackage[numbers]{natbib}
\usepackage[usenames,dvipsnames]{color}
\usepackage{soul}
\usepackage[resetlabels,labeled]{multibib}              
\newcites{PS}{Final Set of Selected Primary Studies}    
\usepackage{lscape}
\usepackage{longtable}
\usepackage{subfiles}

\begin{document}

\title{Immersive Learning Frameworks:\\A Systematic Literature Review}

\author{Filipe Arantes Fernandes, Claudia Susie Camargo Rodrigues, Eldânae Nogueira Teixeira and Cláudia Werner}

\markboth{UNPUBLISHED}
{Fernandes \MakeLowercase{\textit{et al.}}: Immersive Learning Frameworks: A Systematic Literature Review}

\IEEEpubid{\makebox[\columnwidth]{0000--0000/00\$00.00~\copyright{}2022 IEEE \hfill} \hspace{\columnsep}\makebox[\columnwidth]{ }}

\maketitle

\begin{abstract}
\textit{Contribution}: This secondary study examines the literature on immersive learning frameworks and reviews their state of the art. Frameworks have been categorized according to their purpose. In addition, the elements that compose them were also categorized. Some gaps were identified and proposed as a research roadmap.

\textit{Background}: Immersive technologies for education have been used for some years. Despite this, there are few works that aim to support the development and use of virtual environments for immersive learning.

\textit{Research Questions}: This systematic review has the following main research question: What is the state of the art of immersive learning frameworks? In order to answer this question, secondary research questions were defined: 1) what definitions of immersive learning were adopted in primary studies? 2) what are the purposes of use by the frameworks? 3) what are the elements that compose the frameworks?
4) what are the methods used to validate the frameworks?

\textit{Methodology}: As per the systematic review guidelines, this study followed a rigorous and replicable process for collecting and analyzing data. From 1721 articles identified in the search engines, 15 were selected after the inclusion and exclusion criteria. 

\textit{Findings}: Most frameworks are models that investigate the causal relationship between immersive learning factors that influence learning outcomes. Although this theoretical aspect is important for the advancement of research, the area still lacks more practical frameworks that address more technical details and support development, as well as the use of immersive virtual environments by teachers and instructors.

\end{abstract}


\section{Introduction} \label{sec:introduction}

\IEEEPARstart{I}{n} October 2021, Facebook, now called Meta\footnote{https://about.fb.com/news/2021/10/facebook-company-is-now-meta/}, announced the plan to create the metaverse, a kind of universe in Augmented and Virtual Reality (AVR). After this announcement, the possibility of using immersive technologies for business, marketing, games and education became even more popular \cite{narin2021content}.

Immersive technologies refers to computer systems (hardware and software) that enable a more intuitive human-computer interface through devices and sensors that interact with up to the 5 human senses. The main devices are Head-Mounted Display (HMD), also known as a headset, and interaction devices. In addition to hardware, Virtual Environment (VE), also known as a virtual world, it is a three-dimensional computer-generated space where users interact with each other (through avatars) or with other virtual objects \cite{biocca2013communication}. From the point of view of involving real and virtual world elements, \citeauthor{milgram1994taxonomy} \cite{milgram1994taxonomy} classifies applications into Virtual Reality (VR), Augmented Reality (AR), Augmented Virtuality (AV) and Mixed Reality (MR). The eXtended Reality (XR) is an umbrella term that encompasses the entire spectrum of Milgram's continuum \cite{stateofxrreport2021}.

Immersive technologies for education have been used for several years, mainly because virtual environments give the user the feeling of being present in the context that is presented, in addition to allowing the virtual manipulation of objects \cite{fialho1999knowledge}. In situations where being physically present would be too expensive, dangerous or impossible, immersive experiences bring many advantages, for example in the training of surgical skills, pilots and astronauts \cite{fialho1999knowledge}. Furthermore, immersive experiences have greater engagement and allow greater interactivity of the student with the instructional material, encourages the collaborative construction of knowledge, presents more contextualized tasks, less abstract instructions and favors reflective practice \cite{lee2014learning}. To the specific use of immersive technologies to improve learning outcomes, the term Immersive Learning (iL) is known to define this research scope.

Although virtual environments have already evolved a lot, there are still many research challenges involving immersive technologies in education \cite{RW49,ghinea2014mulsemedia}. In addition to the complexity of generating computational solutions for the specifics of educational demands, generally developed by researchers in the field of computing, there are challenges regarding pedagogical and psychological aspects, as well as user experience, storytelling, simulator sickness and others. In parallel, devices have evolved rapidly, allowing the use of virtual environments both in traditional devices (desktop and mobile) and in immersive devices (head-mounted display, motion sensors, and others). Despite the range of devices, developing for the various platforms is also challenging in order to ensure effective platform-independent performance. These and other main challenges for the adoption of immersive technologies in education are grouped into 6 categories, according to the State of XR Report \cite{stateofxrreport2021}: access, affordability, inadequate XR teacher training programs, interoperability, lack of content and lack of infrastructure and tech support.


In order to solve the challenges mentioned above and to contribute to the evolution of research related to immersive technologies in education, iL Frameworks have been a strategy that researchers have found to define a basic conceptual framework to gather concepts and design a comprehensive understanding of a given phenomenon in the context of iL \cite{regoniel2015conceptual}. Therefore, considering the challenges of immersive technologies in education and frameworks as solutions to these problems, this systematic literature review aims to provide evidence on the state of the art of iL Frameworks. More precisely, we are interested in understanding what the purposes are, the elements that compose them and how the frameworks contribute to the solution of the main challenges, according to the State of XR Report \cite{stateofxrreport2021}, in addition to identifying gaps and opportunities for future research.

This article is organized as follows: 
Section \ref{sec:related} presents some previous secondary studies on Immersive Learning. Section \ref{sec:method} describes the research method and the article selection process. Section \ref{sec:results} presents the answers to the research questions. Section \ref{section:discussion} discusses the relevant findings, as well as a research roadmap and, finally, conclusions and future work are presented in Section \ref{sec:conclusion}.

\section{Related Work}\label{sec:related}
In order to systematize the selection of related works, a search for studies was carried out through a simplified review protocol. In January 2022, we ran the following search string (adapted from the tertiary study of \citeauthor{kitchenham2009systematic} \cite{kitchenham2009systematic}): \textit{(TITLE-ABS-KEY(``immersive learning" OR ``immersive education") AND TITLE-ABS-KEY(``review of studies" OR ``structured review" OR ``systematic review" OR ``literature review" OR ``literature analysis" OR ``in-depth survey" OR ``literature survey" OR ``meta-analysis" OR ``past studies" OR ``subject matter expert" OR ``analysis of research" OR ``empirical body of knowledge" OR ``overview of existing research" OR ``body of published research")) AND (EXCLUDE(DOCTYPE,``cr"))}. Only the Scopus search engine was used, as it indexes a variety of digital libraries. Furthermore, it is not our focus to rigorously perform the selection of other secondary studies related to this one. As a result of the search, 16 documents were returned, 7 of which were secondary studies in iL, which will be described in the following.

\citeauthor{wu2020effectiveness} \cite{wu2020effectiveness} and \citeauthor{snelson2020educational} \cite{snelson2020educational} investigated learning performance through immersive technologies. More specifically, \citeauthor{wu2020effectiveness} \cite{wu2020effectiveness} compared the effects of immersive VR with non-immersive VR. As a result, the 35 studies analyzed indicated that immersive VR is more effective than non-immersive VR. In addition, they also identified that immersive VR has a great impact on K-12 learners; STEM (Science, Technology, Engineering and Mathematics) and in the development of specific skills and in the simulation of real situations. On the other hand, \citeauthor{snelson2020educational} \cite{snelson2020educational} focuses on applications that use low-cost equipment through 360º videos. The authors investigated how 360º videos are used and what are their advantages and disadvantages in education.

\citeauthor{huang2021systematic} \cite{huang2021systematic} performed a systematic review in order to find primary studies that report the use of AVR for language teaching. We found 88 articles published in 2011 and 2020, which were analyzed from the following perspectives: tools used, student profile, main findings, reason why virtual learning environment are used and their implications. The study mainly concludes that AVR raises the level of learning; university students are the main users of immersive technologies and the benefits found are improved learning outcomes and increased motivation.

\citeauthor{ntaba2021open} \cite{ntaba2021open} focus on how immersive technologies can support distance learning. More precisely, they investigated what the challenges are and how AVR is used to support distance learning.

\citeauthor{qiao2021integration} \cite{qiao2021integration} and \citeauthor{rey2021effectiveness} \cite{rey2021effectiveness} focus on training. \citeauthor{qiao2021integration} \cite{qiao2021integration} investigated the effectiveness of immersive virtual reality simulation in interprofessional education. Among the 12 primary studies selected, it was concluded that immersive technologies value the approach of shared and team learning. \citeauthor{rey2021effectiveness} \cite{rey2021effectiveness} synthesized outcome criteria to measure the effectiveness of work at heights training with VR in various contexts. From the 21 documents analyzed, the results support safety managers and practitioners, providing a catalog of training methods, effects and assessment indicators.

Finally, \citeauthor{morgado2020unifying} \cite{morgado2020unifying} performed a review of secondary studies and produced a literature review protocol specifically for the scope of iL.

In general, the works above sought evidence of improvement in learning outcomes after intervention with immersive technologies. Each study focused on a context, application domain and immersive technology type. Our study differs from the others, as we are interested in obtaining the state of the art of frameworks that support the advancement of iL research, being cause and effect models of variables that influence learning, as well as guidelines to support the practice of developing immersive educational environments and recommendations for use by educators and students. Therefore, we consider the absence of a systematic review on iL frameworks as a gap in the literature that must be filled.

\color{black}
\section{Research Method}\label{sec:method}
The research method of this secondary study follows three main phases of a systematic literature review proposed by \citet{kitchenham2007guidelines}. The first phase is associated with planning the review, in which the protocol is developed and evaluated. Once the protocol is defined and validated by the researchers involved, it begins the phase of conducting the review, in which the objective is to select the primary studies, extract and perform the data synthesis. Finally, the last phase defines the mechanisms for the dissemination of the results found with the study. The review process is detailed below.
\subsection{Research Questions}
In this study, the following main research question was defined: \textit{what is the state of the art of iL frameworks?} 
A framework is understood as a supporting structure that aim to guide the achievement of iL objectives. In order to answer this main question, secondary research questions were defined:
\begin{itemize}
\item RQ1: What definitions of iL were adopted in primary studies?
\item RQ2: What are the purposes of use by the frameworks?
\item RQ3: What are the elements that compose the frameworks?
\item RQ4: What are the methods used to validate the frameworks?
\end{itemize}

The purpose of RQ1 is to identify the meaning of iL used by the authors, since its definition is not consolidated by the technical literature. RQ2 aims to understand how frameworks support the use of XR in teaching and learning, for example, frameworks support the development of immersive applications or the use of virtual environments, such as Second Life. One of the main contributions of this review is related to RQ3. Immersion, sense of presence and flow, among others, are common terms in this area, but they have ambiguous definitions. For example, Slater and Mel \cite{slater2003note} state that immersion is related to the characteristics of immersive devices, while \citet{jennett2008measuring} define that it is associated with cognitive issues. In this way, this research question aims to identify the main elements that compose each framework, as well as the meaning of the concepts and theoretical background that contributed to the design of the frameworks. Finally, RQ4 has assessed the purpose of understanding how the frameworks were assessed.

\subsection{Search Process}

It was established as a search process the construction of a search string that automatically returns articles in the Scopus, IEEE Xplore, ACM Digital Library, Science Direct and Web of Science databases. In order to support the definition of the search string, a set of terms was established following the PIO paradigm \cite{kitchenham2007guidelines}:
\begin{itemize}
\item \textit{Population}: immersive education, immersive learning, immersive teaching, immersive training;
\item \textit{Intervention}: augmented reality, mixed reality, virtual reality, extended reality;
\item \textit{Outcome}: framework, design, guideline, model.
\end{itemize}

The “OR” boolean operator was used to join the related terms and the “AND” boolean operator to join the terms of population, intervention and outcome. In addition, “NOT” boolean operator was used as a filter strategy to avoid articles on artificial intelligence without the context of human learning \cite{radianti2020systematic}. In this way, the search string is defined as: \textit{(“immersive education” OR “immersive learning” OR “immersive teaching” OR “immersive training”) AND (“augmented reality” OR “AR” OR “mixed reality” OR “MR” OR “virtual reality” OR “VR” OR “extended reality” OR “XR”) AND (framework OR design OR guideline OR model) AND NOT (“artificial intelligence” OR “deep learning” OR “machine learning” OR “neural network”)}.

The search string has been validated in the Scopus database to be able to return the following control articles: [PS\citenum{dengel2020immersive}, PS\citenum{gupta2019design}, PS\citenum{ip2018design}, PS\citenum{klippel2020immersive}, PS\citenum{schott2018virtual}].

These control articles were defined by four reviewers: one professor and researcher with large experience in experimental software engineering; two postdoctoral researchers; and one doctoral student. All reviewers are interested in immersive technologies in software engineering education. After this validation, the search for the articles started. 
\subsection{Selection Criteria and Procedure}
This section describes the conduction of the review phase. In November 2021, the search string was executed in the title, abstract and keywords metadata for each database. In the end, 1721 results were obtained: ACM (127), IEEE Xplore (841), Science Direct (163), Scopus (277) and Web of Science (313). In order to start the selection procedure, the following exclusion criteria were applied by reviewers while reading title, abstract and keywords:
\begin{itemize}
\item EC1: Duplicate article;
\item EC2: Article not being a primary study;
\item EC3: Article being a work in progress or short paper;
\item EC4: Article not published in journal, conference or book chapter;
\item EC5: Authors having a most recent article;
\item EC6: Article not reporting as main contribution generic method or approach that supports the development or selection of immersive educational applications.
\end{itemize}

After applying these criteria, 28 studies were eligible for full text reading and the following inclusion criteria were applied:
\begin{itemize}
\item IC1: Article being accessible for download;
\item IC2: Full text article written in the English language;
\item IC3: Article answers at least one research question from the review.
\end{itemize}

As a result, 12 articles were selected. 
During the reading of each article, three steps were performed at the same time: data extraction; quality assessment and snowballing. For each article, the one-level backward snowballing technique \cite{wohlin2014guidelines} was carried out in order to identify other studies potentially relevant for this secondary study through bibliography references. The first two steps were applied for each article selected by the snowballing. At end, 3 studies were added to this review.

\begin{figure}[!t]
\centering
\includegraphics[width=3.25in]{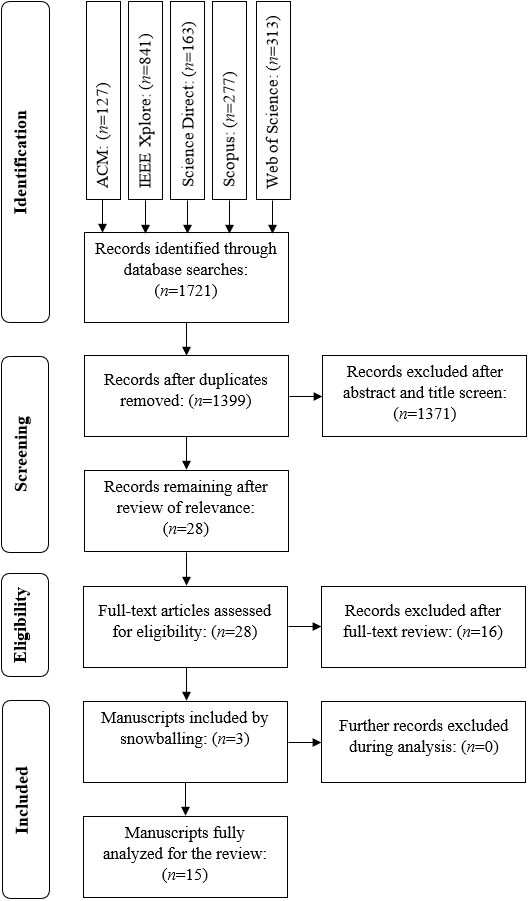}
\caption{Overview of the filtering process.}
\label{fig:prisma}
\end{figure}

After the inclusion criteria and snowballing, 15 articles were selected to compose the final set of articles for this secondary study. Fig. \ref{fig:prisma} shows all the steps taken to find the final set of articles. The organization of the steps was inspired by the PRISMA method \cite{moher2009preferred}.

An electronic spreadsheet was used to support the data extraction process as well as the quality assessment. The quality of selected articles was evaluated according to the questions: 
\begin{itemize}
\item QA1: How clear was the framework's purpose?
\item QA2: How well was the way of using the framework described?
\item QA3: How well were the framework elements described?
\item QA4: How well was the framework validation performed?
\end{itemize}

All researchers reviewed each article's score, according to the following scale (one value per question): 0 - poorly; 0.5 - reasonably; 1 - well. Considering this score, 2 studies reached 4 points, 3 studies reached 3.5 points, 3 studies reached 3 points, 2 studies reached 2.5 points, 3 studies reached 2 points and 2 studies reached 1 point. Regarding quality questions, QA1 was attended by 84\% of studies, QA2 by 47\% of studies, QA3 by 75\% of studies, and QA4 by 63\% of studies. 
Despite the low score, we decided to maintain the studies because we have identified several gaps that can produce interesting discussions and insights for future research, mainly from the perspective of using and validating the frameworks.

Briefly, Table \ref{tab:selection-process} shows the number of studies selected through the sources, studies excluded according to the inclusion and exclusion criteria and, finally, the studies selected for data extraction.
\begin{table}[!h]
\begin{center}
\caption{Studies Selected and Included}
\label{tab:selection-process}
\begin{tabular}{ l c c c c }
\hline
Source & \#papers & \#papers  & \#final \\
& selected & excluded & papers\\
\hline
ACM & 127 & 127 & 0\\
IEEE Xplore & 841 & 837 & 4\\
Science Direct & 163 & 163 & 0\\
Scopus & 277 & 270 & 7\\
Web of Science & 313 & 312 & 1\\
Snowballing & 3 & 0 & 3\\
\hline
Total & 1724 & 1709 & 15\\
\hline 
\end{tabular}
\end{center}
\end{table}

\subsection{Threats to Validity}

Despite the contribution of this study, we identified some threats to validity. The analysis is based on the 15 selected articles. For a secondary study, this number of articles can compromise the results. iL is a recent area of research and this fact may explain the amount of studies. Another factor that should also be considered is the use of the term ``immersive learning" and its variations in the search string. A search with related terms, such as virtual learning worlds, iL experiences, and others, could lead to a large volume of articles that would be out of scope. One of the main reasons for keeping the focus of our study was to obtain works that respond at least to RQ1. Furthermore, the research method was based on systematic literature review guidelines \cite{kitchenham2007guidelines} to ensure the quality of this study.

Out of 15 articles, 4 did not validate the approach and this factor can be considered a threat to validity. These articles were kept with the aim of obtaining the maximum amount of studies and achieving a more assertive overview of the area. Furthermore, even though they did not meet some defined quality criteria, they are studies published in journals and conferences and reviewed by the scientific community peers.

\section{Results}\label{sec:results}
The previous section we presented how the 15 primary studies were selected, that is, what sources were used, search string defined, inclusion and exclusion criteria and other details of the selection process. In this section, we will answer each research question based on the data extracted from the primary studies. 

\subsection{What definitions of iL were adopted in primary studies (RQ1)?}
Immersive technologies as support to education has been used by scientific community for decades. There is a range of studies that adopted augmented and virtual reality as an improving mechanism of the learning outcomes in many areas of knowledge \cite{bacca2014augmented}. Unfortunately, the lack of consensus on concept definition is not something new in augmented and virtual reality. Many researchers have divergent definitions about immersion and presence. From this scenario, we would like to know what are the iL definitions used in studies. 

From 15 selected studies, 5 defined iL with two points of view. Firstly, the following authors believe that iL is related to, mainly, pedagogical and subjective aspects. In the case of [PS\citenum{dengel2020immersive}], did not define it directly, but we understand that iL is the achievement of learning outcomes through educational virtual environments. Therefore, the authors established variables (immersion, presence and learning potential) influence the learning outcomes. In the study [PS\citenum{abdelaziz2014immersive}] believes that the iL concept supports self-regulated, self-determined, self-controlled, informal and life-long learning through a cognitive engagement network that starts with the student and goes through the pro-action engagement, acting engagement, reflection engagement, and reaction phases.

On the other hand, iL is defined considering technological aspects [PS\citenum{klippel2020immersive}, PS\citenum{cardona2019architectural}]. iL is immersive experiences for place-based education [PS\citenum{klippel2020immersive}]. In other words, it is to support the learning through immersive virtual field trips. According to [PS\citenum{cardona2019architectural}], the users must achieve their learning objectives through a transfer of iL based on virtual reality to the real world with real situations through hands-on activities, interacting with objects and events in the simulated world.

A definition that is between the two points of view above is used by [PS\citenum{ip2018design}]: iL is to use technologies, especially computer graphics and human-computer interaction technologies, to create simulated virtual worlds, in which learning can take place by employing appropriate instructional and pedagogical approaches. The authors consider technological and pedagogical aspects. 

We believe that understanding the definition of iL is very important for the advancement of future research. Through the findings, we realized that there is no consensus about what is iL. Clearly there is a separation between pedagogical and technological aspects. Although [PS\citenum{dengel2020immersive}] consider the educational virtual environments, they highlight that immersion, presence and learning potential are main variables to achieve iL. Moreover, [PS\citenum{abdelaziz2014immersive}] focused on an approach based on the constructivist model. Only [PS\citenum{ip2018design}] highlighted the importance of pedagogical and technological aspects.

In addition to the definition given by the authors, we identified two papers [PS\citenum{dengel2020immersive}, PS\citenum{koutromanos2021mobile}] published in International Conference of the iL Research Network (iLRN) \cite{iLRN}. This conference aims to connect researchers, educators and developers in order discuss how XR can provide various opportunities for education.  Thus, we also consider iL as a recent research area. 

Therefore, in our point of view, iL could be defined as a research area that investigates how to improve the learning outcomes through the relationship between the triad immersive technologies, psychological and pedagogical aspects. Considering the main elements extracted from each framework (see Section \ref{section:discussion}) these three aspects were confirmed.

\subsection{What are the purposes of use by the frameworks (RQ2)?} \label{section:rq2}
In this research question, we want to understand for what purposes the iL frameworks were proposed.

In general, from the point of view of the objective, the works were categorized into theoretical and practical. Theoretical frameworks are models that establish the relationship between factors that influence learning outcomes or the adoption of immersive technologies, as well as elements that support the design of learning activities in immersive educational environments. On the other hand, we consider work that establishes guidelines or development models that support the production of immersive educational environments as practical frameworks.

In addition to this broad categorization between theoretical and practical framework, we created subcategories to establish a better understanding regarding the contribution of each work. Table \ref{tab:theoretical-frames} shows the classification of theoretical frameworks and Table \ref{tab:practical-frames} of practical frameworks.

About theoretical frameworks, the works [PS\citenum{abdelaziz2014immersive}, PS\citenum{de2010learning}] were classified as \textit{design of learning activities}, because they define elements that must be considered to design learning activities and assess whether immersive educational environments fit the specificities of the activities. More specifically, [PS\citenum{abdelaziz2014immersive}] aims to be immersive Web-based learning model for supporting learning through phases that virtual worlds should provide to students to achieve learning, while [PS\citenum{de2010learning}] aims to be an evaluation methodology for designing learning activities in virtual worlds as well as evaluating learning experiences. 

\textit{Factors that influence learning outcomes} category classifies works that define elements that are relate to and influence learning outcomes. In general, the works model a causal relationship of the main elements that each author considers important in iL to explain the influence of learning outcomes through immersive educational environments. These elements for some authors are denominated affordances [PS\citenum{schott2018virtual}, PS\citenum{chan2019affordances}, PS\citenum{dalgarno2010learning}, PS\citenum{fowler2015virtual}], objective and subjective factors [PS\citenum{dengel2020immersive}] and variables [PS\citenum{klippel2020immersive}, PS\citenum{lee2010does}].

In the \textit{factors  that  influences teachers’ intention} category the model proposed by [PS\citenum{koutromanos2021mobile}] determines the teachers’ intention to use Augmented Reality applications. This work, based on Technology Acceptance Model (TAM) \cite{tam1989}, helps to understand what are the main characteristics that applications must have to comply with educational purposes, from the teacher's point of view.

Finally, regarding practical frameworks, [PS\citenum{aguayo2020framework}, PS\citenum{misbhauddin2018vredu}] define a set of \textit{guidelines} and design principles for immersive environment development for educational purposes. Specifically, [PS\citenum{aguayo2020framework}] consider design principles and processes that can enhance learning outcomes within free-choice settings, such as museums and visitor centres and [PS\citenum{misbhauddin2018vredu}] developed a general framework that transports all elements of the classroom (from instructor’s point-of-view) to the immersive virtual environment.

Lastly, in the \textit{development model} category includes the works that minimally define a development process to be followed (steps), the actors involved (students, instructors and developers), the types of immersive technologies, as well as software design tools [PS\citenum{gupta2019design}, PS\citenum{ip2018design}, PS\citenum{cardona2019architectural}].

\begin{table}
\begin{center}
\caption{Classification of Theoretical Frameworks}
\label{tab:theoretical-frames}
\begin{tabular}{ l l }
\hline
Subcategories & Primary Studies \\
\hline 
Design of learning activities & [PS\citenum{abdelaziz2014immersive}, PS\citenum{de2010learning}]\\ 
Factors that influences learning outcomes & [PS\citenum{dengel2020immersive}, PS\citenum{klippel2020immersive}, PS\citenum{schott2018virtual}, PS\citenum{chan2019affordances}, \\
& PS\citenum{dalgarno2010learning}, PS\citenum{fowler2015virtual}, PS\citenum{lee2010does}]\\ 
Factors that influences teachers' intention & [PS\citenum{koutromanos2021mobile}]\\
\hline
\end{tabular}
\end{center}
\end{table}

\begin{table}
\begin{center}
\caption{Classification of Practical Frameworks}
\label{tab:practical-frames}
\begin{tabular}{ l l }
\hline
Subcategories & Primay Studies \\ 
\hline
Guideline & [PS\citenum{aguayo2020framework}, PS\citenum{misbhauddin2018vredu}]\\ 
Development model & [PS\citenum{gupta2019design}, PS\citenum{ip2018design}, PS\citenum{cardona2019architectural}]\\ 
\hline
\end{tabular}
\end{center}
\end{table}

\subsection{What are the elements that compose the frameworks (RQ3)?}
Once the frameworks have been categorized according to their objective as per the previous section, in this research question we are interested in understanding what are the main elements that composes each framework.

During data extraction, the elements of each framework were categorized in order to group similar aspects among them and, mainly, to assist in data analysis. This categorization emerged during the reading of the works, considering the reviewers' experience in the areas of immersive technology and education. When it comes to immersive technology, two fundamental aspects must be considered when developing an application: technological (devices, infrastructure, platform etc.) and psychological (the feeling of being present in the virtual world, emotions, awareness etc.). In education, the main aspect considered in this work context was the pedagogical. Therefore, the frameworks elements were grouped into technological, psychological and pedagogical aspects, in addition to the combination between them. All tabulated data considered for this analysis can be accessed at the electronic address
\url{http://reuse.cos.ufrj.br/ilframeworks/}.


\begin{figure}[!t]
\centering
\includegraphics[width=3.25in]{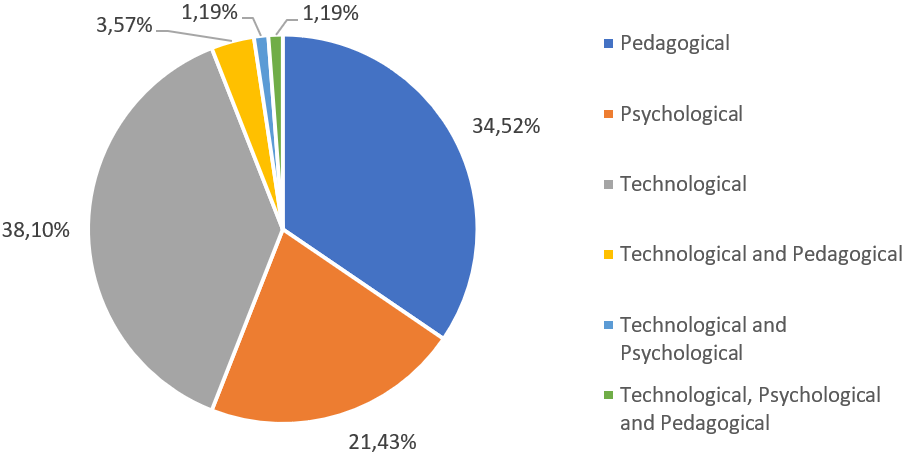}
\caption{Overview of the framework elements.}
\label{fig:chart-rq3-pizza}
\end{figure}

Fig. \ref{fig:chart-rq3-pizza} presents an overview of the result of categorizing the frameworks elements. Therefore, 38.10\% of the works consider technological aspects, 34.52\% consider pedagogical aspects and 21.43\% psychological aspects. In some cases, elements addressed more than one aspect. Thus, 3.57\% represent technological and pedagogical aspects, 1.19\% represent technological and psychological and, finally, 1.19\% are related to technological, psychological and pedagogical aspects.

Through this analysis, in most of the works, the concepts of immersion and sense of presence were approached, but with divergent views. Some of them consider immersion as a technological property and others as a mental state of belonging to the virtual world. In addition, some authors claim that immersion and a sense of presence can be considered as a psychological aspect, that is, a feeling of belonging to the virtual world. These and other findings will be discussed in detail in Section \ref{section:whats-im-pre}.

\subsection{What are the methods used to validate the frameworks (RQ4)?}
Finally, the last research question aims to understand how the frameworks were validated. Out of 15 works, 4 did not validate the proposal. Table \ref{tab:frames-validation} presents a list of the works and the type of validation.

\begin{table}
\begin{center}
\caption{Frameworks Validation}
\label{tab:frames-validation}
\begin{tabular}{ l l }
\hline
Categories & Primay Studies \\ 
\hline
Action research & [PS\citenum{klippel2020immersive}, PS\citenum{aguayo2020framework}]\\ 
Post-test & [PS\citenum{abdelaziz2014immersive}]\\
Pre- and post-test & [PS\citenum{dengel2020immersive}, PS\citenum{schott2018virtual}, PS\citenum{cardona2019architectural}, PS\citenum{chan2019affordances}, PS\citenum{koutromanos2021mobile}, PS\citenum{lee2010does}]\\
Pre- and post-test and & [PS\citenum{ip2018design}, PS\citenum{de2010learning}]\\
observation of student performance &\\
\hline
\end{tabular}
\end{center}
\end{table}

Most studies adopted the validation strategy through \textit{pre- and post-tests}. Participants answer a questionnaire (pre-test) to record their knowledge before the intervention, as well as obtain demographic data. Afterwards, the participants perform some tasks and, at the end, answer another questionnaire (post-test). Some works chose to add \textit{observation of students performance} to the pre- and post-test. [PS\citenum{de2010learning}] used video observations of the real world and the in-world sessions as well as recordings and chat logs and [PS\citenum{ip2018design}] lexical analysis of the learners' comments.

The works [PS\citenum{klippel2020immersive}, PS\citenum{aguayo2020framework}] used the strategy that can be classified as \textit{action research}. Specifically, [PS\citenum{klippel2020immersive}] carried out a set of evaluations and the results were used to evolve their proposal and [PS\citenum{aguayo2020framework}] defined their guidelines based on feedback during the development of immersive applications. Finally, [PS\citenum{abdelaziz2014immersive}] selected a group of students and divided them into a control group and an experimental group. Each group was selected according to the already known profile and at the end of the tasks a \textit{post-test} questionnaire was used.

\section{Discussion} \label{section:discussion}
In this section we discuss the main findings, how frameworks solve the main barriers to adoption and also propose some issues to include in the research roadmap on iL frameworks.

\subsection{What is immersion and presence?}\label{section:whats-im-pre}

One of the main characteristics of Virtual and Augmented Reality (AVR) is to provide to the user the feeling of belonging to the virtual environment. Real-time interaction and the human senses stimulated by the device's sensors enhance the feeling of presence in a virtual environment. Immersion and presence are concepts found in frameworks and their definitions are antagonistic, that is, some authors have divergent understandings about these concepts.

For example, [PS\citenum{cardona2019architectural}, PS\citenum{dalgarno2010learning}, PS\citenum{dengel2020immersive}, PS\citenum{klippel2020immersive}, PS\citenum{lee2010does}] understand that immersion corresponds to the properties and capabilities of technology to stimulate the human sensory system. All these works were based on Slater's works \cite{slater2003note,slater1997framework,slater1999measuring}. From this point of view, immersion is a quantifiable description of the technology, that is, one must consider the quality of the graphic display, stereo audio, haptic sensor, motion sensor, among other characteristics of the device.

By focusing in detail on this idea of immersion, the authors are concerned with establishing which technological capabilities the devices provide in order that experiences can match expectations of interaction with the virtual environment.

Considering the Milgram's continuum \cite{milgram1994taxonomy}, for the user to can experience the sensation of fully belonging to the virtual environment, devices that stimulate all human senses are needed. The most common senses provided by XR headsets, are sight and hearing. Oculus Quest 2, HTC Vive Pro 2 and Microsoft HoloLens 2 are examples of a range of devices of this type, in addition to allowing interaction with the virtual environment through controls for each hand. Some devices have limitations, such as Google Cardboard, in which the user's interaction with the virtual environment is through a fixed pointer on the screen, denominated gaze point. Thus, to interact with some object in the virtual environment, the user must position the pointer with the movement of the head and wait for the timer (a few milliseconds) until the interaction is completed.

XR headsets have evolved over the years and have become more accessible to the public due to reduced cost and improved technology. There are devices that stimulate other human senses and have more intuitive interactivity, but some of these require high investment for acquisition or are still prototypes. Omni One is a system produced by Virtuix in which it captures human body movements such as walking, running, jumping and rotating the body, allowing the user to have a greater immersive experience when moving naturally in the virtual environment. Another example of an interaction device is Leap Motion, which allows the user to interact with the virtual environment by capturing hand movements. 

Another human sense that can be explored is smell. ION\footnote{https://ovrtechnology.com/technology/} is a device produced by OVR Technology that allows the user to smell it during the immersive experience. In addition to smell, prototypes are being developed to allow the user to feel the taste, as is the case with the work of Karunanayaka et al. (2018) \cite{karunanayaka2018new}. Kim et al. (2020) developed a prototype to sense the temperature of virtual objects \cite{kim2020thermal}. Finally, HaptX Gloves DK2\footnote{https://haptx.com/} is a haptic device which allows the user to feel the contact and weight of virtual goals when interacting with them.

The works [PS\citenum{schott2018virtual}, PS\citenum{chan2019affordances}] understand that immersion can be defined as a mental state in which the user is surrounded by another reality demanding his/her attention. This definition of immersion is similar to the concept of presence, also called by some authors as a sense of presence. All works that define the concept of presence are unanimous in stating that it is a user's mental state of belonging to the virtual environment in which they are interacting. For [PS\citenum{fowler2015virtual}], immersion goes beyond technological and psychological points of view. In his framework, the learning experience is also considered, called pedagogical immersion, in which it is the pedagogical state that arises from learning in a virtual environment.

The above works explicitly defined the understanding of immersion and presence according to the application context. Table \ref{tab:im-pre-concepts} presents the definitions of each framework, as well as the reference used for each concept. Although some works have not explicitly defined immersion and presence, the meaning of these concepts can be understood throughout the reading of the full text, as in the case of [PS\citenum{ip2018design}, PS\citenum{abdelaziz2014immersive}, PS\citenum{de2010learning}] in which the understanding is that immersion is a mental state, contrary to [PS\citenum{gupta2019design}, PS\citenum{aguayo2020framework}, PS\citenum{misbhauddin2018vredu}], who understand immersion as a technological aspect. Exceptionally, [PS\citenum{koutromanos2021mobile}] do not address these concepts.

\begin{table*}[ht]
    \caption{Immersion and Presence Concepts}
    \label{tab:im-pre-concepts}
    \centering
    \begin{tabular}{
        p{0.1\linewidth}
        p{0.05\linewidth}
        p{0.5\linewidth}
        p{0.1\linewidth}
    }
    \hline
    \textbf{Concepts}	&	\textbf{Primary studies}	&	\textbf{Definitions}	&	\textbf{Based on}\\ 
    \hline
        
Immersion	&	[PS\citenum{dengel2020immersive}]	&	Quantifiable description of technology	&	\cite{slater2003note}	\\
	&	[PS\citenum{klippel2020immersive}]	&	It refers to system characteristics	&	\cite{slater1999measuring}	\\
	&	[PS\citenum{schott2018virtual}]	&	A state where the user (the learner in this context) is surrounded with another reality claiming their complete attention	&	\cite{murray1997hamlet}	\\
	&	[PS\citenum{cardona2019architectural}]	&	Immersion technology aspects are considered to offer the user the feeling of presence in an artificial environment as if he were in a daily learning situation	&	\cite{lavalle2016virtual}	\\
	&	[PS\citenum{chan2019affordances}]	&	The mental state of total absorption in the virtual environment enabled by, in addition to a high degree of real-time interaction, the rich information perceived through multiple sensory channels	&	\cite{burdea2003virtual}	\\
	&	[PS\citenum{dalgarno2010learning}]	&	Immersion relies on the technical capabilities of VR technology to render sensory stimuli	&	\cite{slater1999measuring,slater2003note,slater2004colorful}	\\
	&	[PS\citenum{fowler2015virtual}]	&	A concept that can bridge both the technological, psychological and pedagogical experiences	&	Authors	\\
	&	[PS\citenum{lee2010does}]	&	Techonological properties	&	Authors	\\
\hline Presence	&	[PS\citenum{dengel2020immersive}]	&	Perception of non-mediation	&	\cite{lombard1997heart}	\\
	&	[PS\citenum{klippel2020immersive}]	&	Mental state	&	\cite{slater1997framework}	\\
	&	[PS\citenum{schott2018virtual}]	&	The subjective experience of being in one place or environment, even when one is physically situated in another	&	\cite{witmer1998measuring}	\\
	&	[PS\citenum{chan2019affordances}]	&	Equal to immersion	&	Authors	\\
	&	[PS\citenum{dalgarno2010learning}]	&	Presence or sense of presense is context dependent and draws on the individual’s subjective psychological response to VR	&	\cite{slater1999measuring,slater2003note,slater2004colorful}	\\
	&	[PS\citenum{fowler2015virtual}]	&	It is the psychological state that can arise from an immersive system	&	Authors	\\
	&	[PS\citenum{lee2010does}]	&	It is the psychological sense of ``being there" in the environment generated by the system	&	Authors	\\
	&		&	Presence is a human reaction to immersion	&	\cite{slater2003note}	\\
	&		&	Presence refers to the user’s subjective psychological response to a system	&	\cite{bowman2007virtual}	\\
	&		&	The sense of presence in a 3-D environment occurs as a consequence of the fidelity of representation and the high degree of interaction or user control, rather than just a unique attribute of the environment	&	\cite{dalgarno2002contribution}	\\

     \hline
    \end{tabular}
\end{table*}

Through the discussions above, it can be seen that the concept of presence is well defined, while immersion has several understandings. In order to understand the relationship between the definitions of the concepts with the objective of each framework, Table \ref{tab:immersion-mapping} presents the mapping between the framework categories, as well as the subcategories, with the types of immersion. Most of the works that consider psychological immersion are theoretical frameworks and the works that consider technological immersion are equally grouped between theoretical and practical frameworks. In particular, all theoretical frameworks that consider technological immersion are concentrated in the \textit{influences learning outcomes} category. Only [PS\citenum{fowler2015virtual}] defined immersion as a set of technological, psychological and pedagogical aspects.

\begin{table*}[ht]
    \caption{Immersion Mapping}
    \label{tab:immersion-mapping}
    \centering
    \begin{tabular}{p{0.1\linewidth}p{0.175\linewidth}p{0.1\linewidth}p{0.1\linewidth}p{0.15\linewidth}}
        \hline
        \textbf{Category}  & \textbf{Subcategory}  & \textbf{Psychological Immersion}  & \textbf{Technological Immersion} & \textbf{Technological psychological and pedagogical Immersion}\\ 
        \hline
        Practical Framework & Guideline & & [PS\citenum{aguayo2020framework}, PS\citenum{misbhauddin2018vredu}] &  \\ 
     & Model Development & [PS\citenum{ip2018design}] & [PS\citenum{gupta2019design}, PS\citenum{cardona2019architectural}] &  \\ 
    \hline
    Theoretical Framework & Design of Learning Activities  & [PS\citenum{abdelaziz2014immersive}, 	PS\citenum{de2010learning}] & &  \\
    & Influences Learning Outcomes & [PS\citenum{chan2019affordances}] & [PS\citenum{dengel2020immersive}, 	PS\citenum{klippel2020immersive}, 	PS\citenum{dalgarno2010learning}, PS\citenum{lee2010does}] & [PS\citenum{fowler2015virtual}] \\ 
     & Influences Teacher's Intention & [PS\citenum{schott2018virtual}] & &  \\
     \hline
    \end{tabular}
\end{table*}

Through the above analysis, we observe that the concept of immersion will vary according to the purpose of each framework. For example, the models of [PS\citenum{dengel2020immersive}, PS\citenum{klippel2020immersive}, PS\citenum{dalgarno2010learning}, PS\citenum{lee2010does}] establish which variables influence learning outcomes in virtual environments. In order to isolate the characteristics of the devices with their ability to ``immerse" the user in the virtual environment, the term immersion was defined as a technological aspect and presence as a psychological aspect of belonging to the virtual environment. From this point of view, an immersive device is not a guarantee to provide the user with a complete sense of presence, because it depends on other factors, such as the proper functioning of the interaction between the user and the virtual environment, motion sickness, fidelity in the graphical representation, interference from the environment external, among others. The works [PS\citenum{ip2018design}, 	PS\citenum{schott2018virtual}, 	PS\citenum{abdelaziz2014immersive}, 	PS\citenum{chan2019affordances}, PS\citenum{de2010learning}] prioritize other variables and consider immersion as a psychological state of belonging to the virtual environment (psychological immersion).

In our view, immersion should be considered as a technological aspect and presence as a psychological aspect of belonging to the virtual environment. In this way, we believe to facilitate the understanding of these trivial concepts and support the identification of the potential of devices and virtual environments to ``immerse" the user and transmit a sense of presence. Therefore, the greater the involvement of the human senses together with the human-computer interaction intuitive, the greater the degree of immersion and potentially the user will achieve the sense of presence. For example, Oculus Quest 2 has the greatest potential to provide presence compared to Google Cardboard, meaning the former is more immersive than the latter, however the reach of the sense of presence depends on several factors throughout the immersive experience.

Therefore, we conclude that immersion must be considered as an objective aspect that characterizes the technological capacity to evoke the user's feeling of presence in a virtual environment and presence a subjective aspect in which the user believes ``being there" in the virtual environment that he/she is interacting with.

\subsection{Finding Solutions to Barriers to Adoption}
As presented in Section \ref{sec:introduction}, one of the objectives of this work is to verify how the frameworks solve the main challenges pointed out in the State of XR Report \cite{stateofxrreport2021}. This report is a body of knowledge based on research-based evidence on “what works” in iL. Organized by the Immersive Learning Research Network (iLRN), a nonprofit organization that connects researchers and educators, experts grouped the main barriers to adoption of XR into:
\begin{itemize}
\item \textit{Access (B1)}: it addresses issues related to limiting the distribution of immersive technologies and the improvement of technologies and applications for people with disabilities.
\item \textit{Affordability (B2)}: economic availability is still a challenge for the implementation of immersive technologies, since the equipment has a high cost;
\item \textit{Inadequate XR Teacher Training Programs (B3)}: both technological and pedagogical support are incipient or non-existent and educators need training programs on topics related to immersive technologies, as well as instructional design support to successfully integrate XR into teaching practices;
\item \textit{Interoperability (B4)}: much immersive content is still locked into certain hardware, software and commercial frameworks. VEs for each XR headset creates an effective environment for educators within which they are then limited, even though the same device can provide many benefits for education;
\item \textit{Lack of Content (B5)}: as with all emerging technology, educators face the challenge of finding immersive instructional content or reusable content to leverage for educational use;
\item \textit{Lack of Infrastructure and Tech Support (B6)}: 
immersive applications depend on the integration between hardware, software and network in order to meet educational purposes. Therefore, ensuring access to immersive experiences considering the available infrastructure and the audience context is still a challenge for iL.

\end{itemize}


As presented above, the State of XR Report \cite{stateofxrreport2021} describes the main challenges for the adoption of immersive technologies in education. In this way, we are interested in finding out if the frameworks help, in some way, in solving the main challenges identified by the report. In this sense, we developed questions that correspond to the barriers to adoption in order to support the mapping:
\begin{itemize}
\item Do frameworks consider aspects of audience disability (B1)?
\item Do frameworks consider the economic availability for the feasibility of immersive experiences (B2)?
\item Do frameworks consider aspects of technical and pedagogical support to institutions and educators (B3)?
\item Do frameworks consider interoperability aspects between applications and devices (B4)?
\item Do frameworks consider aspects of resource reuse (B5)?
\item Do frameworks consider infrastructure aspects (B6)?
\end{itemize}

By rigorously analyzing the data extracted from the frameworks, we found that no work directly addresses the above questions. Considering our classification of works, we expected some response from practical frameworks rather than theoretical ones. This can be explained because adoption barriers correspond to technological and practical aspects rather than theories and pedagogical approaches.

As can be seen in Figure \ref{fig:chart-rq3-pizza}, the elements of each framework are analyzed and categorized. We observed that practical frameworks address generic or context-specific issues. For example, \citeauthor{ip2018design} \cite{ip2018design} showed a methodology for supporting the design iL experiences to MOOC learners through iterative stages and \citeauthor{aguayo2020framework} \cite{aguayo2020framework} proposed a set of design principles and guidelines for self-determined mixed reality learning. Both jobs are domain-specific. On the other hand, \citeauthor{gupta2019design} \cite{gupta2019design} adopted Information-Centric Systems Engineering (ICSE) principles to guide the development of immersive technologies, but did not consider specific aspects of XR.

In this sense, when reflecting on this critical point of frameworks and considering the background of the researchers, we present below a research roadmap that defines important issues about iL frameworks.

\subsection{Research Roadmap}
In this review, we identify frameworks with specific purposes, some of which model the causal relationship between factors that can influence learning outcomes, methods for designing activities in virtual environments, as well as guidelines and models that define the main requirements of virtual educational environments.

Understanding what and how factors influence learning outcomes is important to ensure that the teaching and learning process is successful. Through theoretical frameworks one is able to understand that there are several factors involved in iL, such as sense of presence, educational context, learning strategies, pedagogical theories and others \cite{dengel2020immersive,klippel2020immersive,schott2018virtual}. Pedagogical and psychological aspects are fundamental in iL. The practical frameworks identified in this review are limited, as their contributions are context-specific and it is not possible to generalize. In general, frameworks define ``what'' and not ``how'' to develop applications for iL. More specifically, the works do not detail nuances of immersive technology in order to guide the development of virtual environments from the point of view of frameworks that address the details of technology to support development, and more practical approaches that support teacher and instructor decision-making.

As regards the frameworks analyzed in this secondary study, in the identified gaps, as well as the experience of reviewers in iL and Software Engineering, we list below some aspects that are fundamental to support the development for iL, which can be considered as research roadmap.

\subsubsection{Level of Immersion} \label{sec:level-imm}
Since human beings have five senses (smell, taste, sight, hearing and touch) to interact with the world they live in, researchers have sought to make users interact with virtual environments in the same way as they interact with the real world, making the immersive experience more complete. Thus, it is important to define which human senses will be involved during the iL experience and which will be the forms of interaction with the virtual environment. This decision will directly influence the choice of immersive devices.

\subsubsection{Immersive Devices}
We consider traditional devices as multimedia (desktop, tablet and smartphone) and mulsemedia \cite{ghinea2014mulsemedia} as immersive devices that raise the level of multimedia immersion and add other human senses such as smell, taste and touch, in addition to providing more natural and intuitive interactions. Examples of mulsemedia devices are XR headsets, haptics, motion sensors and others. Each immersive device has characteristics that will influence the experience as a whole. For example, the immersive experience via smartphone is more limited compared to the XR headset. At the same time, the associated cost (B2) must also be considered, as pointed out by the State of XR Report \cite{stateofxrreport2021}. Considering devices that meet desired immersion levels and affordability is a challenge that must be taken into account to meet audience requirements.

\subsubsection{Development Tools}
Developing for XR is complex, because it needs a multidisciplinary team that involves skills such as coding, game design, 3D modeling, storytelling, user experience and others. For each specialty a set of tools is needed to produce the artifacts. For example, to create 3D objects and scenarios, it is necessary to master 3D modeling tools such as Blender\footnote{https://www.blender.org/}, 3DS Max\footnote{https://www.autodesk.com/ products/3ds-max/} and Maya\footnote{https://www.autodesk.com/products/maya/}. Software engineers are not required to master these tools, nor have the ability to model 3D objects, but they must be able to specify the trivial development tools related to the chosen platforms. Therefore, priority must be given to which platform the application will run on, that is, whether it will be a native or web application.

Then the development environment to implement the virtual environment features must be chosen. Most immersive device manufacturers provide the Software Development Kit (SDK) according to development environments such as Android, iOS, Web, Unity\footnote{https://unity.com/}, Unreal\footnote {https://www.unrealengine.com/} and others.

Therefore, if Google Carboard will be used in an immersive experience and the virtual environment must be downloaded to the smartphone (native), the developer must choose the development environment that they are corresponding familiar with (Android NDK, iOS or Unity) and import the SDK. If the virtual environment is run via browser, it is necessary to choose a tool compatible with the parameters of the virtual environment so that the immersive experience is not impaired. Examples of web frameworks are WebXR\footnote{https://immersive-web.github.io/}, A-Frame\footnote{https://aframe.io/}, Babylon.js\footnote{https://www .babylonjs.com/} and React 360.

Identifying technologies and developing applications that are interoperable (B4) is another step towards mitigating problems related to adoption barriers.

\subsubsection{User Experience}
Unlike applications based on the WIMP interface (Windows, Icons, Menus and Pointers), immersive applications need attention to avoid uncomfortable experiences. Instructions on how to interact with the virtual environment must be accessible at all times. Oculus Quest 2 Controller, for example, has 6 buttons for each hand and this can be a lot of information for the user. Therefore, the environment must provide a training section so that the user can gradually get used to the virtual environment. In addition, the virtual environment must maintain a stable frames per second (fps) rate, preferably 60 fps, to keep camera movement in the environment corresponding to the user's head movement, avoiding discomfort. For this, a series of restrictions in the development is recommended, such as limiting the amount of polygons; using just one camera instead of \textit{post-processing effects} for \textit{draw calls} needed in the scene; use of panoramic images (360 degrees) and others.

\subsubsection{Simulator Sickness}
Simulator sickness is a very important aspect in XR and one that no iL framework has addressed. Users may experience uncomfortable symptoms (such as eyestrain, fatigue, dizziness, ataxia) that make the immersive experience difficult. Regan and Ramsey (1994) \cite{regan1994frequency} found that individuals exposed to the virtual environment had symptoms for up to 5 hours after the experience. The severity and duration of these symptoms can be influenced by the time of exposure to the virtual environment and the intensity of the experience \cite{kim2005characteristic}. Thus, the way the user will move in the virtual environment is an important precaution to avoid discomfort during and after the experience. In this example, implementing the teleportation technique reduces the probability of the user presenting motion sickness symptoms instead of walking freely through the virtual environment.

\subsubsection{Accessibility Technologies}
As presented in the State of XR Report \cite{stateofxrreport2021}, one of the issues preventing the adoption of immersive technologies is the inadequacy of applications and devices for people with disabilities (B1). WalkinVR\footnote{https://www.walkinvrdriver.com/} is the first major app that allows for adjustments in XR experiences based on users' height and various disabilities. However, more research and new applications are critically needed to address these issues and involving both academia and industry is paramount.

\subsubsection{Experience Reuse}
From the point of view of Software Engineering, software reuse is an approach that starts from the principle of enhancing the use of existing software, aiming to reduce production and maintenance costs, guarantee more agile deliveries, try to add more quality and maximize the return on investment of software \cite{werner1997memphis}.

Following this line, Domain Engineering (DE) and Application Engineering (AE) can be applied as techniques to improve the development of immersive applications. DE is the process of identifying and organizing knowledge about a class of problems, the problem domain, to support its description and solution \cite{prieto1990domain}. For example, there are domains (e.g. STEM, health and military education) that have common characteristics and that, therefore, their applications could be built from the same process and artifacts, thus promoting the reuse of common concepts and functionalities.

While DE is concerned with developing artifacts for reuse, AE builds applications based on the reuse of artifacts and models generated by DE. According to \citeauthor{northrop2007framework} \cite{northrop2007framework}, AE develops software products based on the artifacts generated by the DE process.

Therefore, the adoption of software reuse techniques as a development strategy has the potential to allow the reuse of assets involved in immersive experiences (B5), as well as improving the quality of applications.

Below, we list some aspects to support teachers and instructors in adopting immersive teaching experiences.

\subsubsection{Immersive Platforms}

iL supports teaching in any field of knowledge. Therefore, teachers and instructors do not have the skills to develop applications and therefore need tools to support their classes. There is a range of platforms that provide immersive content.

Talespin's training platform\footnote{https://www.talespin.com/} puts the user directly into a guided scenario, in a realistic two-person discussion situation. Engage VR\footnote{https://engagevr.io/} and Unimersiv\footnote{https://unimersiv.com/} are online training and education platforms that have immersive content from various areas of knowledge.

In addition, teachers and instructors can create virtual spaces and insert their contents. Frame VR\footnote{https://framevr.io/} and Mozilla Hubs\footnote{https://hubs.mozilla.com/} are examples of current web tools that allow one to create virtual classrooms where students can access simultaneously through avatars, communicate and interact with each other. Furthermore, Second Life\footnote{https://secondlife.com/} and Open Wonderland\footnote{http://openwonderland.org/} are virtual spaces that have been much explored by the scientific community.

Therefore, these immersive platforms must be prepared to support educators (B3). These professionals do not have technical skills and need intuitive tools to support the adequacy of instructional content to immersive technologies.

\subsubsection{Available Infrastructure}
Using immersive experiences in teaching still emerges as a challenge, as immersive equipment still requires considerable investment. Therefore, the choice of device that will be used in the teaching and learning process directly impacts the pedagogical performance. Desktops and smartphones are more common devices among people than XR headsets. On the other hand, the educational institution can choose to purchase immersive devices, but it will require a high investment.

Another point is the hardware configuration and Internet connection speed (B6). Virtual environments demand high performance in image processing and take up a lot of storage space. In the case of a web application, the connection quality will also impact the performance of learning activities.

\subsubsection{Improved Learning Outcomes}
This is one of the great challenges in iL. Ensuring improved learning outcomes through immersive experiences still requires further empirical studies. Increasing the degree of immersion of devices is not a guarantee of high academic performance. The sense of presence and pedagogical aspects (quality of instructional content and pedagogical theories) are also important and must be considered to achieve effectiveness in learning outcomes. Therefore, further studies are needed on how to improve learning considering technological, psychological and pedagogical aspects.

\subsubsection{Learning Analytics}
Through the analysis carried out in this study, it was identified that no framework considered the monitoring of student learning in virtual environments. In general, virtual environments immerse students in instructional content, but learning data are not captured to analyze students' educational performance. We believe that Learning Analystics can provide valuable information about actual performance and improve the teaching-learning process. For example, through biofeedack sensors, heart rate, breathing, sweat, and temperature readings can indicate whether the student in a particular section felt any discomfort \cite{reis2021learning}.

\section{Conclusion}\label{sec:conclusion}
This systematic literature review aimed to identify the state of the art of immersive learning frameworks. Through the 15 selected articles it was possible to obtain an overview of the contributions and identify gaps and research opportunities.

Through the research questions, we identified that the authors have divergent understandings about immersive learning (RQ1), as well as the definition of immersion. In addition, we grouped the works regarding the purpose of use and we found that there are frameworks to support the design of learning activities, identify which factors influence the learning and intention of teachers, guidelines and development models (RQ2). We also found that frameworks are composed of three main aspects: technological, psychological and pedagogical (RQ3). Finally, most of the frameworks were validated through a questionnaire, but 4 articles did not validate the proposal (RQ4).

The relevance of this study lies in the discussion and definition of the concept of immersion, better understanding of immersive learning, identification of gaps and the proposal of a research roadmap so that frameworks can address the development of immersive environments greater detail, as well as the use of experiences immersive by teachers and instructors.

As future works, the results and discussions of this systematic literature review can be used to guide research and propose frameworks that aim to address implementation details and, consequently, obtain more effective virtual environments and corroborate immersive learning. \citet{fernandes2021work} is an example of a framework that supports the planning of the development of immersive educational applications, considering the characteristics of technologies, skills, competences and pedagogical approaches in the context of Software Engineering Education.

\section*{Acknowledgments}
The authors would like to thank CNPq and CAPES Brazilian Agencies for financial  support.

\bibliographystyle{IEEEtranN-ital}
{\footnotesize\bibliography{bibliography}}

\bibliographystylePS{IEEEtranN-ital}
{\footnotesize\bibliographyPS{primarystudies}}
\nocitePS{dengel2020immersive}
\nocitePS{gupta2019design}
\nocitePS{ip2018design}
\nocitePS{klippel2020immersive}
\nocitePS{schott2018virtual}
\nocitePS{abdelaziz2014immersive}
\nocitePS{aguayo2020framework}
\nocitePS{cardona2019architectural}
\nocitePS{chan2019affordances}
\nocitePS{dalgarno2010learning}
\nocitePS{fowler2015virtual}
\nocitePS{de2010learning}
\nocitePS{koutromanos2021mobile}
\nocitePS{lee2010does}
\nocitePS{misbhauddin2018vredu}

\newpage

\vfill

\end{document}